# The Perils of Advocacy

Joel Atkins[1]

## Abstract:

Statisticians and data scientists find insights that help lead to better understanding and better outcomes. When clients and managers come to us for help (and even when they don't), we want to share our advice. While we should be free to share our recommendations, we need to be clear about what the data is telling us and what is based "only on our judgment".

Gelman, et. al. wrote "As we have learned from the replication crisis sweeping the biomedical and social sciences, it is frighteningly easy for motivated researchers working in isolation to arrive at favored conclusions—whether inadvertently or intentionally." One senior business leader I know said, "if you have data, great; if we're just going on intuition we can use mine". However, having data isn't enough. We need to be rigorous in our analysis to avoid finding insights that aren't supported. This paper will go through a number of examples to illustrate common mistakes.

## Introduction:

Data scientists have a number of incentives to find results. Whether it's justifying our salaries, the need to publish papers, or supporting a cause we believe in.

Gelman, et. al.[2] wrote "As we have learned from the replication crisis sweeping the biomedical[3] and social sciences[4], it is frighteningly easy for motivated researchers working in isolation to arrive at favored conclusions—whether inadvertently or intentionally. As a partial solution, former federal judge Richard Posner has suggested that courts appoint independent experts to sift through the statistical evidence." They are not discussing fraud, but an often well intentioned effort to continue looking at the data different ways until one finds something interesting or helpful. There is a flip side to this, where researchers miss some necessary steps and find results that are not supported by the data.

Below, I will offer some examples where I think that we have not done the due diligence to support our conclusions. While this happens everywhere, I will draw my examples from the public domain – usually published papers. In this article, I'm not taking a position on any of these issues. I am saying that the analysis was not done well; and I think that should concern all of us on either side of these issues, if we want to preserve our influence.

---

1 The author may be contacted at joelatkins@yahoo.com

2 Gelman, A., Ho, D., Goel, S., (January 14, 2019) What Statistics Can't Tell Us in the Fight over Affirmative Action at Harvard, *Boston Review*, https://www.bostonreview.net/articles/andrew-gelman-sharad-goel-daniel-e-ho-affirmative-action-isnt-problem/

3 Ioannidis, J., (2005) Why Most Published Research Findings Are False, *PLOS Medicine*, https://journals.plos.org/plosmedicine/article?id=10.1371/journal.pmed.0020124

4 Camerer, C.F., Dreber, A., Holzmeister, F. et al. (2018) Evaluating the replicability of social science experiments in Nature and Science between 2010 and 2015. Nat Hum Behav 2, 637–644. https://www.nature.com/articles/s41562-018-0399-z

# Where our data comes from

## Survey data

BMC Infectious Diseases published[5], and later retracted, a paper about the safety of Covid vaccines. This study did an online survey of 2,840 people and asked them 40 questions. These 2,840 people were selected by an opt-in online survey that was balanced by age, sex, and income to be representative of the United States population. 57 of the respondents said that they knew someone who had died after receiving the vaccine and 165 said that they knew someone who had died after having Covid. At the time they wrote their paper, the CDC was estimating 839,993 deaths from Covid. Applying this ratio of 57/165 to 839,993, the authors estimated that there had been 289,789 fatalities from the vaccine. They then estimated that a fraction of these were really due to other causes and only 278,000 were due to the vaccine.

I have a lot of sympathy for epidemiologists and survey statisticians. Understanding what is happening in a large population is challenging. Having said that, there a few problems with this analysis:

- Asking about the experience of the respondents' friends is a good way to collect data about a lot more people. It's also useful when we are concerned that people might be less than honest, such as questions about political views, drug use, and sexual behavior. However, it reduces the integrity of the data. Often, we won't know as much about our friends as we know about ourselves.[6]
- The survey asks about the respondents' friends. Some people have wide social circles and thus have a relatively large chance of being in a respondent's social circle. Others have narrower social circles and thus have a relatively small chance of being included in a respondent's social circle. People in more social circles may have different fatality rates than people in fewer social circles.
- While they did match the US population distribution by age, sex, and income, it is a big assumption that people taking on line surveys are the same as people who do not take on line surveys. This bias shows up both in who sees the survey and who then chooses to take the survey. They had 216 people opt out, not huge compared to the 2,840 respondents, but we also can't assume that they would give the same answers. (This is a challenge for most surveys.)

It is hard to make any inferences from this study. This paper was ultimately retracted[7]. One of the reasons provided was "there are critical issues in the representativeness of the study population and the accuracy of data collection." However, this only happened after it was published and received a lot of publicity[8].

[5] Skidmore, M. (2023) RETRACTED ARTICLE: The role of social circle COVID-19 illness and vaccination experiences in COVID-19 vaccination decisions: an online survey of the United States population. BMC Infect Dis 23, 51. https://doi.org/10.1186/s12879-023-07998-3

[6] The authors do make a small correction based on their estimate of how many people died from heart attacks, strokes, and blood clots. They do not explain how they derived this adjustment and they ignore other possible causes of death. It's possible that many of these people died from Covid.

[7] Retraction Note: BMC Infectious Diseases (2023) 23:51, https://doi.org/10.1186/s12879-023-07998-3

## Survival bias

In any observational study, we are limited to the data we observe, with all of its idiosyncrasies. One challenge with observational data is survival bias. A simple example is with mutual funds. If we only consider funds where we have prices now and five years ago, we will exclude any fund that closed during the last five years. These funds have performed worse than the funds that stayed open, so excluding them will bias our average return higher. One study found that survival bias increased "equal-weighted US domestic equity mutual fund market from 1993 through 2006" by 157 basis points. This was the difference between mutual funds doing better than the passive benchmark (on a risk adjusted basis) and worse than that benchmark.[9]

Survival bias also helps explain the equity risk premium puzzle. The equity risk premium is the incremental return about the risk free rate that investors demand for investing in equities of average risk. This premium has been estimated to be about 6% in the United States. Many are surprised that it is so high, and that question has been named the "equity risk premium puzzle".[10]

In the 1920's, many global economies were in similar shape. Since then, the United States economy has outperformed others for a number of reasons, including the fact that the United States was not heavily impacted by either World War. Other countries, like Argentina, were quite rich, but then declined and stagnated, partly due to political instability. If we look at all the global economies, we see that stocks have not done as well over the last century in other countries and those studying the United States are looking at the economy that did the best during this period. The figure below shows returns that stocks provided from 1920-2020 in seven large countries to illustrate this point.[11] (Poland has a large discontinuity because its market was closed for decades.)

---

[8] Letter on author's website. https://mark-skidmore.com/wp-content/uploads/2023/04/Retraction-Update-for-the-Public.pdf

[9] Rohleder, M., Scholz, H., and Wilkens, M., (September 9, 2010) Survivorship Bias and Mutual Fund Performance: Relevance, Significance, and Methodical Differences, *Review of Finance* 2011, 15 (2), 441-474, https://ssrn.com/abstract=1101615

[10] Damodaran, Aswath, (March 23, 2022) Equity Risk Premiums (ERP): Determinants, Estimation, and Implications – The 2022 Edition. Available at SSRN: https://ssrn.com/abstract=4066060 or http://dx.doi.org/10.2139/ssrn.4066060

[11] van Binsbergen, J., Hua, S., and Wachter, J. (2020) Is The United States A Lucky Survivor: A Hierarchical Bayesian Approach *Wharton School* https://rodneywhitecenter.wharton.upenn.edu/wp-content/uploads/2020/12/30-20.Wachter.VanBinsbergen.pdf

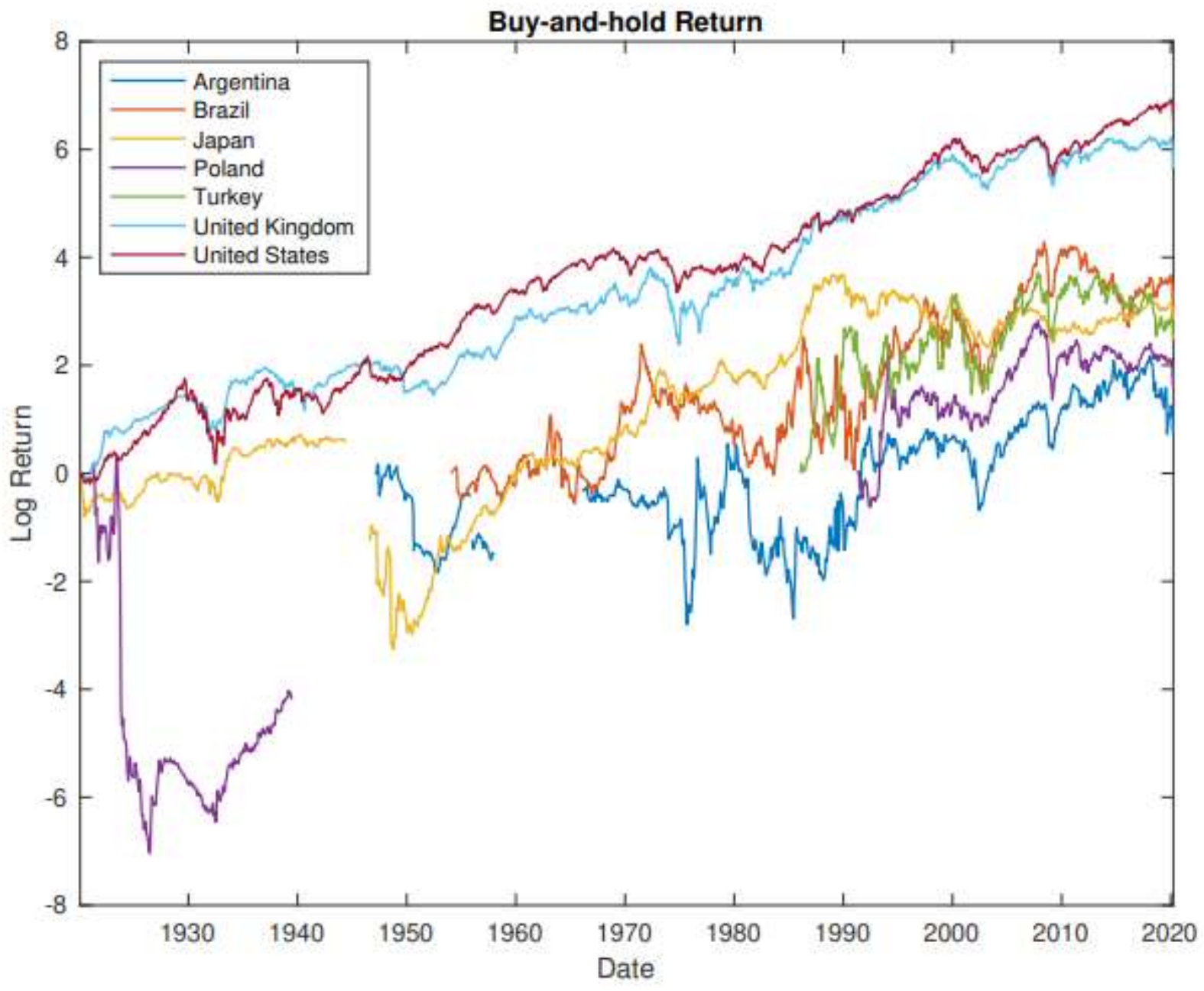

**Figure 3.** Buy-and-Hold Return

The figure plots the buy-and-hold return of investing \$1 in max{1920, when the market enters the sample} for selected countries. For the few countries (e.g. Poland) which had its stock market nationalized, it is assumed that \$1 more is invested at the point of restart. The US stands out in having a smooth and continuous series, whereas other markets experience trading breaks and high volatility.

In general, we should understand what data we are using and what, if any, data has been excluded.

# Using the right data and metrics

In 2017, "Donald Trump signed a second executive order banning people from certain majority-Muslim countries from entering the USA. The previous order banned citizens of seven countries: Syria, Iran, Sudan, Libya, Somalia, Yemen, and Iraq, but was stopped after legal action. The new ban applies to six countries - Iraq is not included - and uses different wording in the hopes of being on firmer legal ground." One data scientists wrote a critique[12] of this ban and submitted it to KDD's Data Science vs Fake News Contest, where it received third place.

The author questions whether this ban would be effective, based on four weak arguments:

- They rely on historical data. This assumes that the environment in the upcoming years will be the same as it was in previous years, something rarely true in geopolitics. One of the articles they link to states "All of those countries are currently roiled by active terror elements such as ISIS and al-Qaeda"[13]. Six months before the Trump proposed this ban, the RAND Corporation wrote about the need to lessen the risk of radicalization of Syrian refugees.[14] In light of these concerns, it is not clear that historical years will give us much useful information.

[12] https://web.archive.org/web/20200508065023/https://www.kdnuggets.com/2017/04/muslim-ban-make-safer-data-science-vs-fake-news.html

- The authors state that after 9/11 (2001) and through 2017, there were only 1.6 people killed per year by foreign born terrorist and none of these terrorist were from the banned countries. This is true, but not that persuasive as successful terrorist attacks are rare and usually small. Excluding or including 9/11 is an arbitrary choice that significantly impacts the number of people killed. Including 9/11 would have added 2,977 additional people[15] who were killed by foreign terrorist. This would increase the number of people killed per year (from 2001 to 2017) from 1.6 to 177. While I don't agree with the decision to use historical data, once the authors chose to go that route, I think they should have included the 9/11 attacks.
- The authors point out that there were only 58 people convicted of terrorism charges banned countries, compared with 130 from other countries and 172 who were from the United States. Thus, the ban would have only affected 16% of convicted terrorists. The raw numbers lack context and are not the right metric. In 2010, roughly the middle of the time period, there were 309M people in the United States[16] of which 40M were born outside the United States[17] including 0.8M from the banned countries[18]. If we looked at rates of arrests per person in the country, we get a much different story.
  - Born in the United States was 172 arrests / (309M total people – 40M not born in the US) = 0.6 arrests per million people.
  - Born outside the United States but not in the banned countries is 130 arrests / (40M born outside the US – 0.8M born in banned countries) is 3.3 arrests per million people.
  - Born in the banned countries is 58 arrests / 0.8M people is 72.5 arrests per million people.

  Thus, we see that the cost of the ban, in terms of the number of people affected, is relatively small compared the number of terrorist who would be affected. This helps us weight the costs and benefits of the ban. This graph shows the difference in using the raw number of people convicted and the rate per million of people. The second column provides a more accurate view and weakens the authors position.

---

[13] David, J. (2017) Trump claims his immigration ban is 'not about religion,' even as order favors Christians, *CNBC*, https://web.archive.org/web/20200508065046/https://www.cnbc.com/2017/01/29/trump-america-proud-nation-of-immigrants-says-this-is-not-about-religion.html

[14] RAND Review (2016) Election 2016: The International Issues. Rand Corporation. https://www.rand.org/blog/rand-review/2016/10/election-2016-the-international-issues.html

[15] https://www.911memorial.org/connect/commemoration

[16] https://www.census.gov/data/tables/time-series/dec/popchange-data-text.html

[17] https://www.migrationpolicy.org/programs/data-hub/charts/immigrants-countries-birth-over-time?width=1000&height=850&iframe=true

[18] https://www.migrationpolicy.org/programs/data-hub/charts/immigrants-countries-birth-over-time?width=1000&height=850&iframe=true

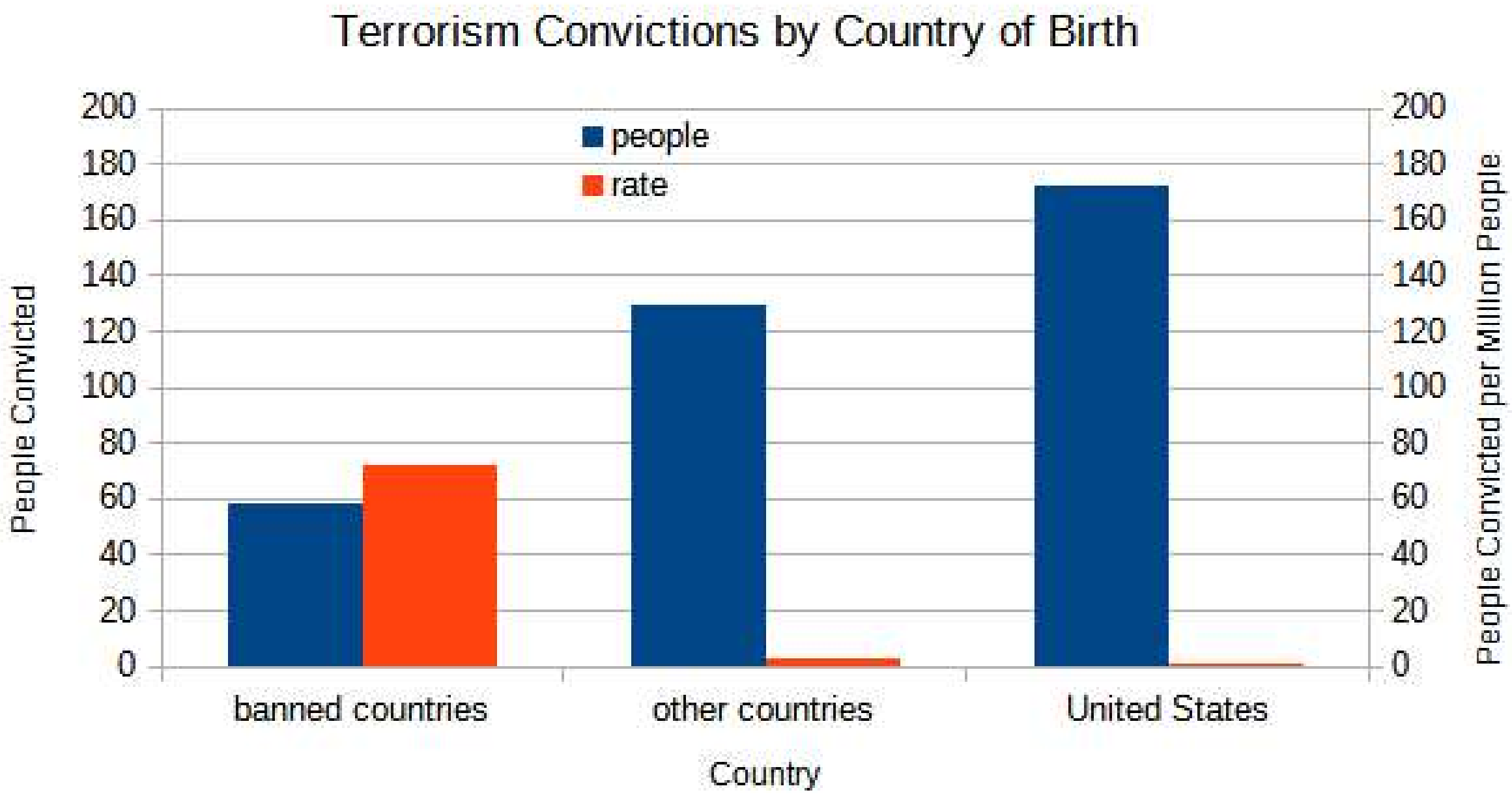

- The author states "However, of those that come from the banned countries, most are naturalized citizens or permanent residents, and thus would not have been affected by the ban (except the first ban, which originally applied to permanent residents)." This is imprecise. Most of the naturalized citizens and permanent residents came to the United States as non-residents, so over time this would eliminate naturalized citizens and permanent residents from the banned countries.

While this case is extreme, it highlights a few important points. Using historical data may not tell us about the future, particularly when the environment is changing quickly. Choosing to use data after 9/11 highlights the importance of what data we choose to use and how we address outliers. Whether we use the number of people arrested or the rate that people are arrested illustrates the importance of how we define metrics. In many of our analysis, we consider ideas that are not well defined. How we measure them is important and can make a difference in our conclusions.

I am not opining on whether the ban was appropriate. There are obvious downsides to the ban which the author mentions. I am saying that the author's arguments were weak.

# Sometimes we try too many things, also known as p-hacking

In the past, researchers had a limited set of potential predictors and a handful of hypotheses, often based on some domain knowledge. They would collect the relevant data and test these hypotheses once the data was compiled. Today, our process can be much more complicated with decision points occurring before we have finished collecting the data. This has led to many more opportunities to get a significant result. Teixeira describes p-hacking as "the exhaustive exploitation of data through the use of different analytical models and/or the manipulation of application criteria of these models until statistically significant results are obtained."[19]

Head, et. al. describe p-hacking as "when researchers try out several statistical analyses and/or data eligibility specifications and then selectively report those that produce significant results. Common practices that lead to p-hacking include: conducting analyses midway through experiments to decide whether to continue collecting data; recording many response variables and deciding which to report postanalysis, deciding whether to include or drop outliers postanalyses, excluding, combining, or splitting treatment groups postanalysis, including or excluding covariates postanalysis, and stopping data exploration if an analysis yields a significant p-value."[20]

In one common type of p-hacking, we find that an effect is not significant, but is tantalizingly close to being significant. Szucs[21] gives several examples of papers where authors have added additional observations to see if the effect is really significant. Stansbury[22] has shown with simulated data that adding observations will give us biased p-values. He ran 10,000 simulations each having ten observations with one predictor and one target that were sampled independently. For the simulations with a p-value between 0.05 and 0.10, he added one more observation. The graph below shows the distribution of p-values with these additional observations. 26% of these borderline predictors now have p-values below 0.05. We see here that adding additional observations greatly increases our Type 1 errors.

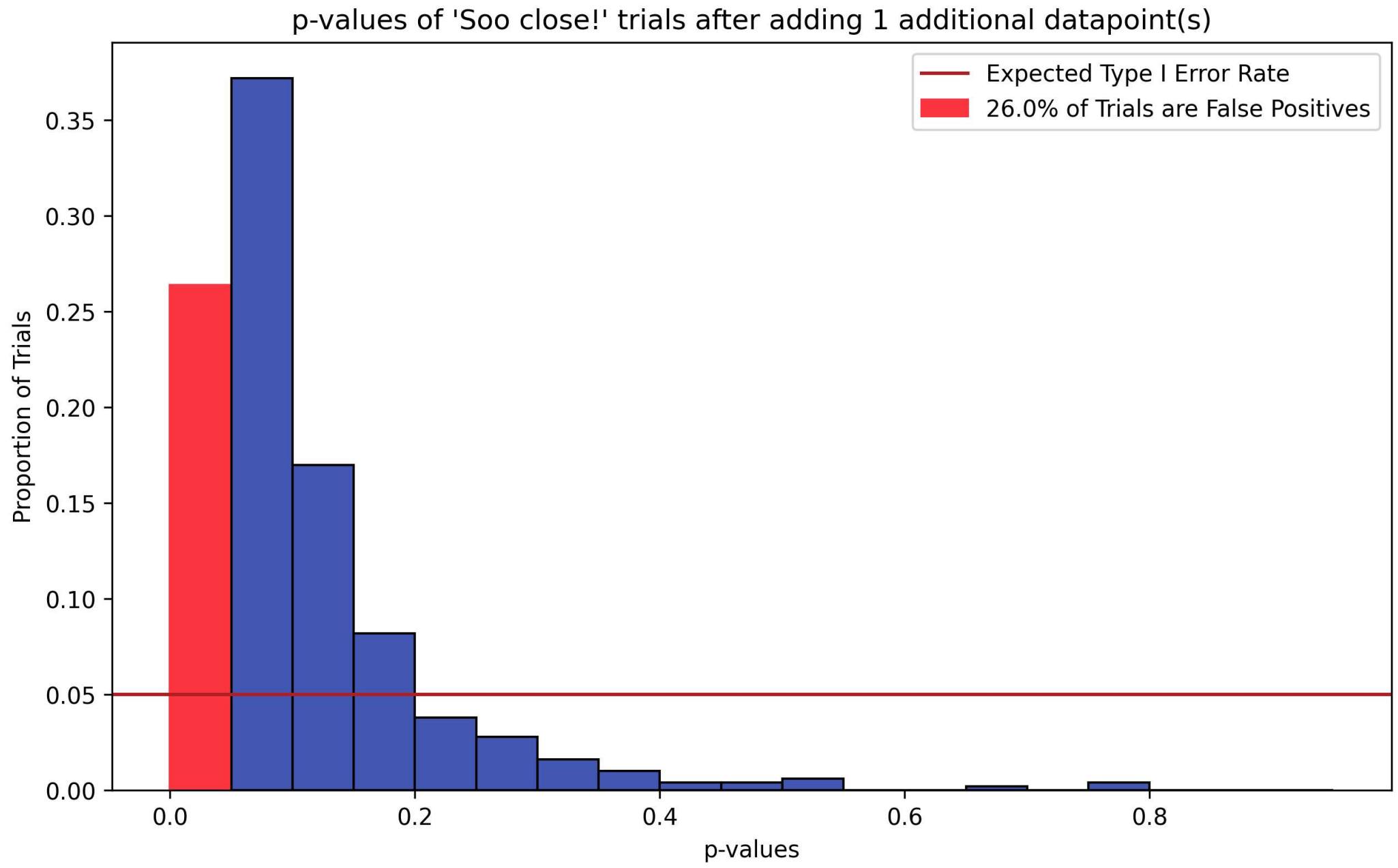

[19] Teixeira, P. (2018) p-Hacking – A call for ethics. *Pulmonology Journal*. https://www.journalpulmonology.org/en-p-hacking-a-call-for-articulo-S2531043718300485

[20] Head, M., Holman, L., Lanfear, R., and Kahn, A. (2015) The Extent and Consequences of P-Hacking in Science. *PLoS Biol.* 13(3). Mar 13 2015. https://doi.org/10.1371%2Fjournal.pbio.1002106

[21] Szucs D (2016) A Tutorial on Hunting Statistical Significance by Chasing N. Front. Psychol. 7:1444. https://www.frontiersin.org/articles/10.3389/fpsyg.2016.01444/full

[22] Stansbury, D. https://dustinstansbury.github.io/theclevermachine/p-hacking-n-chasing

The news site www.fivethirtyeight.com has created an example[23] illustrating other types of p-hacking where we can test whether the economy does better when Democrats are in power and when Republicans are in power.  For both questions, someone can select how do define who is in power, how to measure how well the economy is doing, and whether to exclude recessions.  Based on these decisions, the results vary substantially are sometimes significant for both of the questions.

Gelman and Loken[24] discuss the multiple choices that researchers make and how this can result in exaggerated p-valuesis.  Rubin[25] offers some ways to address it, but they are not simple.  There is also some who fear that Rubin's ideas might stifle science and suggest more training and the idea of "p-diligence".[26]

# Correlations are overrated

In the recent past, physical experiments, data collection, and analysis were more time consuming and expensive.  Today, with much better computers, we can test lots of predictors[27].  If we use a simple p-tests, we will find lots of false positives.  Benjamini and Hochberg did some valuable work[28] to address this, penalizing the tests based on how many predictors are considered.  With this correction we will find fewer false positives, but using a large number of predictors will make it harder to find real effects.

Even if we limit our data to variables that might have some intuitive relation with our target, we can still run into trouble with proxy variables that appear to be good predictors.  Below are two examples of this.

## Picking up proxy variables – example 1

Danzinger, et al.[29] found that parole boards were most likely to grant parole at the beginning of the day or shortly after a break and that the likelihood decreased until the next break.  In an article in Nature[30], Corbyn speculated that this was caused by the judges being tired or hungry.

---

23 https://projects.fivethirtyeight.com/p-hacking/

24 Gelman, A. and Loken, E. (2014) The Statistical Crisis is Science: Data-dependent analysis – a "garden of forking paths" – explains why many statistically significant comparisons don't hold up. American Scientist 102.6:460-465.  https://www.americanscientist.org/article/the-statistical-crisis-in-science

25 Rubin, M. (2017). An evaluation of four solutions to the forking paths problem: Adjusted alpha, preregistration, sensitivity analyses, and abandoning the Neyman-Pearson approach. Review of General Psychology, 21, 321-329. https://doi.org/10.1037/gpr0000135

26 Lombrozo, T. (2014) Science, Trust And Psychology In Crisis, *NPR*. June 2, 2014. https://www.npr.org/sections/13.7/2014/06/02/318212713/science-trust-and-psychology-in-crisis

27 Smith, G. (2020) The paradox of big data. *SN Applied Sciences* 2, 1041. https://doi.org/10.1007/s42452-020-2862-5

28 Benjamini, Y. and Hochberg, Y. (1995) Controlling The False Discovery Rate - A Practical And Powerful Approach To Multiple Testing. *Journal of the Royal Statistical Society. Series B: Methodological.* 57:289-300. https://www.researchgate.net/publication/221995234_Controlling_The_False_Discovery_Rate_-_A_Practical_And_Powerful_Approach_To_Multiple_Testing

29 Danziger, S., Levav, J. & Avnaim-Pesso, L. (2011) *Proceedings of the National Academy of Sciences*. https://www.pnas.org/doi/pdf/10.1073/pnas.1018033108

30 Corbyn, Z. (2011) Hungry judges dispense rough justice. *Nature*. https://www.nature.com/articles/news.2011.227

Scientific America[31] saw this as a case of mental fatigue, confirming the findings of previous studies. This might have been an argument for putting more weight on algorithms in this process.

Weinshall-Margel, et al.[32] found "that case ordering is not random and that several factors contribute to the downward trend in prisoner success between meal breaks. The most important is that the board tries to complete all cases from one prison before it takes a break and to start with another prison after the break. Within each session, unrepresented prisoners usually go last and are less likely to be granted parole than prisoners with attorneys." While Danzinger, et al. ruled out a number of possible explanations they missed these.

We should look for correlated variables that may be explaining what we are seeing. Domain knowledge will help with this, as will a second set of eyes.

# Picking up proxy variables – example 2

Several studies[33,34,35,36] argue that Black drivers are stopped at a higher rates than White drivers. These papers try to control for a number of other variables, but they are still observational studies with limited data. As such, they need to assume that driving behavior does not vary by race and differences in policing are due to some type of bias.

Racial biases in policing is not a new concern. In 1999, the New Jersey Governor and Attorney General accused their highway troopers of racial profiling[37]. The Federal government created a consent decree for the troopers. After this, the troopers reduced their discretionary activity by more than 90%. Even with this reduced activity, Blacks were still being stopped at a disproportionate rate for speeding. At that point, the troopers asked the attorney general to study speeding on the turnpike.

The study[38] commissioned by the Attorney General found that Black drivers were stopped at a rate disproportionate to their share of drivers but in line with the percent of Black drivers who were driving 15 or more miles per hour above the speed limit. This study included a literature review and cited studies saying that:

- In San Diego, 25% of stops are pretext stops that were not related to traffic violations. The police stops there are in line with a 25% weight of criminal suspects demographics and a 75% weight of the population demographics[39].
- In Cincinnati, police stops “appeared to be correlated with driving patterns, crime patterns, drug calls, and overall demand for police services. This suggested that the disproportionate stops of African American drivers may be explained by workload factors rather than biased policing[40].

---

31 Kleiner, K. ( September 1, 2011) Lunchtime Leniency: Judges' Rulings Are Harsher When They Are Hungrier. *Scientific American*. https://www.scientificamerican.com/article/lunchtime-leniency/

32 Weinshall-Margel, K. and Shapard, J. (2011) Overlooked factors in the analysis of parole decisions. *Proceedings of the National Academy of Sciences.* https://www.pnas.org/doi/full/10.1073/pnas.1110910108

33 Davis, E., Whyde, A., Langton, L. (2018) Contacts between Police and the Public, 2015. *Bureau of Justice Statistics*. https://bjs.ojp.gov/library/publications/contacts-between-police-and-public-2015

34 Greenleaf, R., Skogan, W., Lurigio, A., (2008) Traffic Stops in the Pacific Northwest: Competing Hypotheses About Racial Disparity. *Journal of Ethnicity in Criminal Justice*, Vol. 6(I). http://www.skogan.org/files/Traffic_Stops_in_the_Pacific_Northwest.text.pdf,

35 McDevitt, J., Farrell, A., Yee, M. (2003) Providence Traffic Stop Statistics Compliance Final Report. *Northeastern University Institute on Race and Justice.* https://repository.library.northeastern.edu/files/neu:344642/fulltext.pdf

36 Racial Justice Project. (2020) Driving While Black and Latinx: Stops, Fines, Fees, and Unjust Debts. *New York Law School*. https://digitalcommons.nyls.edu/cgi/viewcontent.cgi?article=1007&context=racial_justice_project

37 MacDonald, H. (2002) The Racial Profiling Myth Debunked, *City Journal*, Spring, https://www.city-journal.org/html/racial-profiling-myth-debunked-12244.html

38 Lange, J., Johnson, M., Voas, R., (2005) Testing the Racial Profiling Hypothesis for Seemingly Disparate Traffic Stops. Justice Quarterly: 22(2) https://www.researchgate.net/publication/248967296_Testing_the_racial_profiling_hypothesis_for_seemingly_disparate_traffic_stops_on_the_New_Jersey_Turnpike

The earlier papers assume that the behavior is the same across races and any difference in policing is due to police behavior, but they did not attempt to test or verify it. The Attorney General’s study tested this assumption and the San Diego and Cincinnati papers found better denominators. The last three studies show that the earlier studies’ approach of looking at the rate of police stops and tickets by race is not sufficient to show that any differences are due to police racism.

# Poorly designed studies

The above examples have been from empirical studies. In many ways, things are much easier if we can design our own study. When we do, we need to be careful about other potential issues.

## Ignoring correlated errors

In many experiments, we will run multiple specimens through our treatment at the same time. This may result in correlated errors. If we don’t reflect this in our analysis, we might think our results are more significant than they are. In many experimental fields, including cell biology, there is significant variability between different days and different labs. In light of this, one paper[41] states that “the sample size N used for statistical tests should actually be the number of times an experiment is performed, not the number of cells (or subcellular structures) analyzed across all structures.”

The authors suggest the graphs below to address this issue. Each column considers the same data, but shows it in a different way. In the “NOPE” and “STILL NOPE” columns, they treat all the observations as though they are independent, resulting in minuscule p-values. In the “GOOD” column, they show the mean for each “run” of the experiment, where a run could be a different day or different lab. For this view, the authors consider the means for the six runs of the experiment and find a much more modest p-value. In the “EVEN BETTER” column, they add color coding to each observation to show which of the three runs it is from.

---

[39] Cordner, G., Williams, B., Velasco, A. (2002) Vehicle Stops in San Diego: 2001. https://www.researchgate.net/publication/264875645_Vehicle_Stops_in_San_Diego_2001

[40] Eck, J., Liu, L., Bostaph, L., (2003) Police Vehicle Stops in Cincinnati: July 1-December 31, 2001. https://www.uc.edu/content/dam/uc/ccjr/docs/reports/project_reports/Final_Police_Stops_Report_d6_H.pdf

[41] Lord, S., Velle, K., Mullins, R., Fritz-Laylinb, L., If your P value looks too good to be true, it probably is: Communicating reproducibility and variability in cell biology. https://arxiv.org/ftp/arxiv/papers/1911/1911.03509.pdf

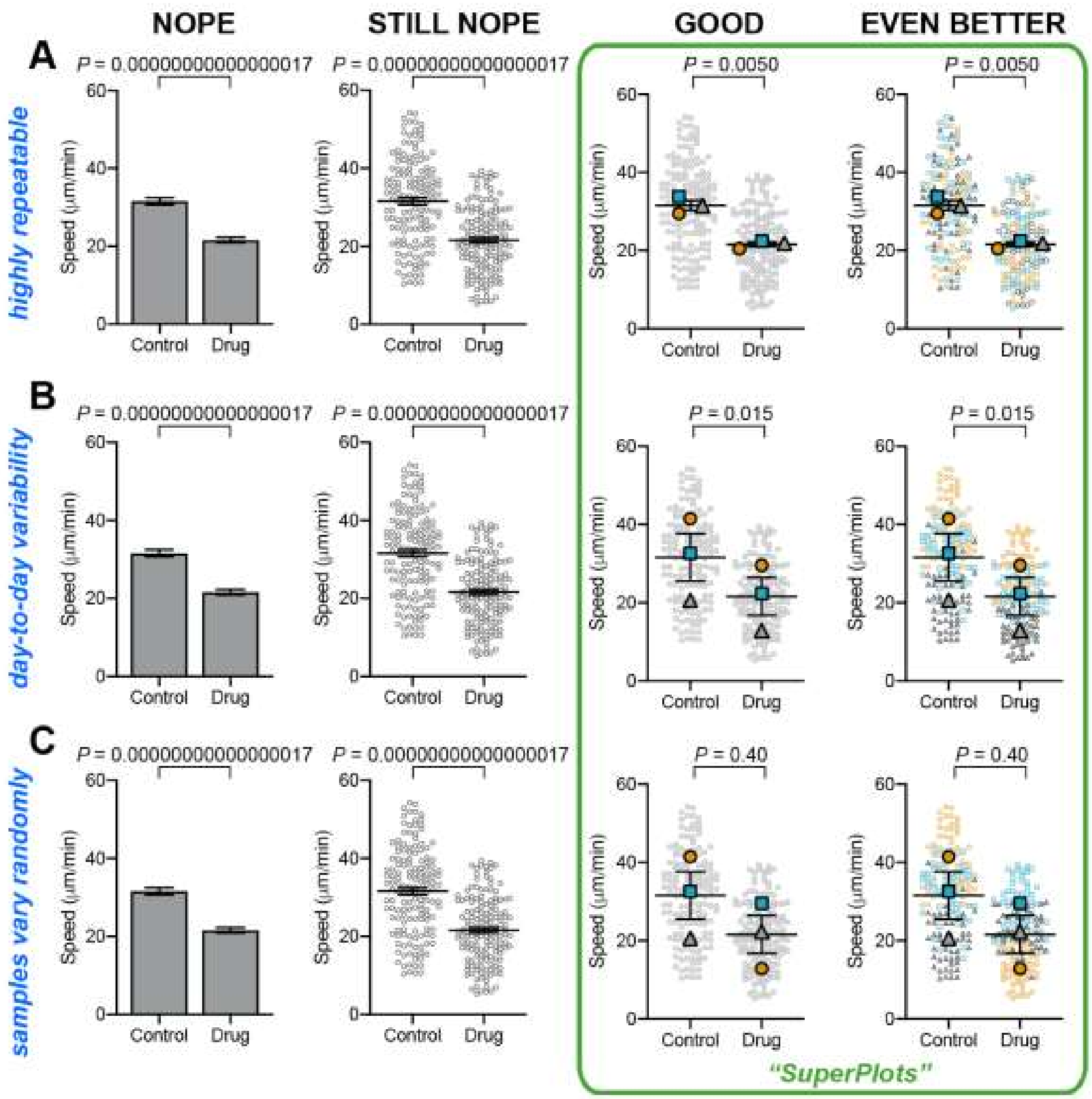

# Selecting an error distribution

Some modelers will transform their target variable. One reason to do this is make the target variable follow a normal distribution. This is based on a belief that "by transforming your target variable, we can hopefully normalize our errors, if they are not already normal."[42] This is not accurate, as normal errors and skewed predictors will give a skewed target variable. This graph[43] shows a case where the target variable is normally distributed conditioned on X (as seen by the sideways black histograms), but not normally distributed (as the blue dots follow roughly a uniform distribution between 40 and 80). This is because the predictors do not follow a normal distribution resulting in the expected value of the target not following a normal distribution.

---

[42] Plummer, A. (2022) Box-Cox Transformation and Target Variable: Explained, *builtin*, https://builtin.com/data-science/box-cox-transformation-target-variable

[43] Li, X, Wong, W, Lamoureux, E, and Wong, T. Are Linear Regression Techniques Appropriate for Analysis When the Dependent (Outcome) Variable Is Not Normally Distributed? *Investigative Ophthalmology & Visual Science* May 2012, Vol.53, 3082-3083. https://doi.org/10.1167/iovs.12-9967

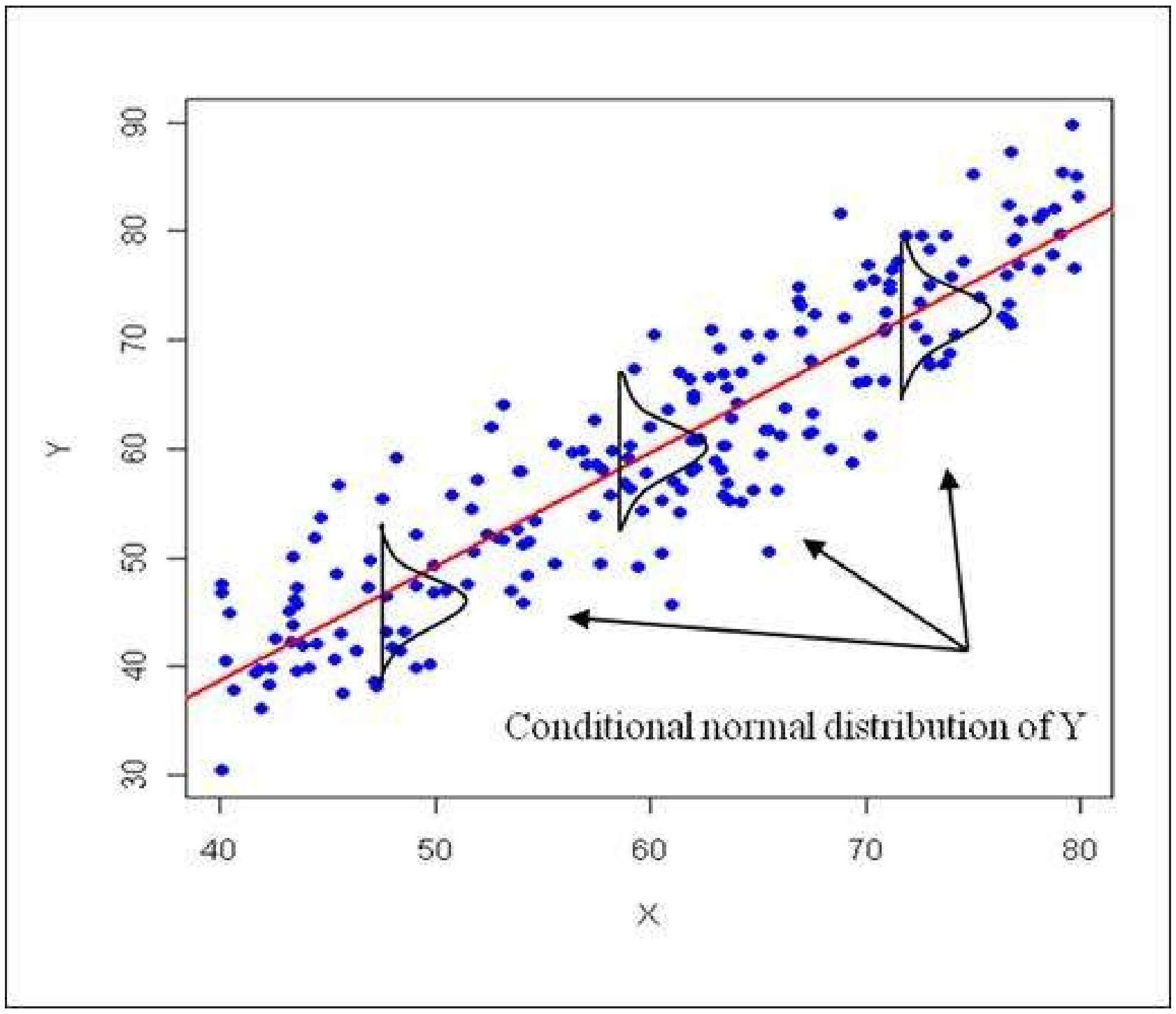

Others transform the target variable because they feel that there is a multiplicative relationship. In this case, the predictors would have a linear relationship with the log of the target variable. However, this approach makes it difficult to describe the actual target variable, as applying the inverse of the original log transformation to the expected value of the log transformed target variable will give an estimate less than the actual expected value of the target[44]. It is better to fit models with a GLM structure and a log-link for the target variable.

## Dimension Reduction

We often have a lot of variables. There are many methods to deal with this. Some, such as principal components, reduce the dimensions of the predictors without considering the target variable. This can lose important information. Consider the following hypothetical example. We look at twenty boys and use their height and weight to predict BMI. This should work well since BMI is weight (kg) / height (m)^2.

Typically here, we would reduce from two dimensions down to one by only keeping the first eigenvector. If we do that in this case, we won't be able to build a useful model. The first eigenvector's large values identify boys who are both tall and heavy - their weights may make sense for their height. The second eigenvector, which we would have removed, does a great job at segmenting - it's under 60 for the boys with the four highest BMIs. It is recognizing boys who are heavy for their weight. By reducing our dimensions without considering our target variable, PCA may remove the information that we need to build a useful model. Options that consider the target will generally give better results[45].

---

[44] More, S. (2022) Identifying and Overcoming Transformation Bias in Forecasting Models https://arxiv.org/pdf/2208.12264.pdf

| age | height (cm) | weight (kg) | BMI | Eigenvector 1 | Eigenvector 2 |
|---|---|---|---|---|---|
| 6 | 103.0 | 14.5 | 13.7 | 109 | 69 |
| 6 | 106.0 | 15.5 | 13.8 | 113 | 69 |
| 6 | 109.0 | 16.5 | 13.9 | 116 | 70 |
| 6 | 100.0 | 17.5 | 17.5 | 107 | 59 |
| 8 | 104.0 | 17.0 | 15.7 | 111 | 64 |
| 8 | 109.0 | 18.0 | 15.2 | 117 | 66 |
| 8 | 114.0 | 20.0 | 15.4 | 122 | 67 |
| 8 | 120.0 | 22.0 | 15.3 | 129 | 68 |
| 10 | 121.0 | 18.0 | 12.3 | 129 | 78 |
| 10 | 127.0 | 21.0 | 13.0 | 136 | 77 |
| 10 | 133.0 | 25.0 | 14.1 | 144 | 74 |
| 10 | 115.0 | 27.0 | 20.4 | 126 | 51 |
| 12 | 128.0 | 23.0 | 14.0 | 138 | 74 |
| 12 | 133.0 | 27.0 | 15.3 | 144 | 69 |
| 12 | 138.0 | 32.0 | 16.8 | 152 | 62 |
| 12 | 143.0 | 36.0 | 17.6 | 158 | 58 |
| 14 | 141.0 | 28.0 | 14.1 | 153 | 75 |
| 14 | 150.0 | 32.0 | 14.2 | 164 | 74 |
| 14 | 158.0 | 37.0 | 14.8 | 174 | 71 |
| 14 | 133.0 | 42.0 | 23.7 | 151 | 34 |

# Anchoring

## Transitory inflation

Sometimes we anchor our expectations to how things were or how we want things to be. During 2021, the Federal Reserve was saying that inflation was transitory. On August 27, 2021, Federal Reserve Chair Jerome Powell said that he wanted to "avoid chasing 'transitory' inflation"[46].On November 30, 2021, Powell said that it was probably a good time to retire the word transitory.[47] During this time, others were warning that:

- In July of 2021, the producer price index was up 7.7% year over year, while the consumer price index was up only 5.3%, suggesting that companies would need to continue raising prices.[48]
- The federal government had a deficits of over $3 trillion in fiscal years 2021.[49]

---

45 Ker-Chau, L. (1991) Slicing Regression: Dimension Reduction via Inverse Regression. *UCLA Department of Statistics*. https://escholarship.org/content/qt3mw1p0mb/qt3mw1p0mb.pdf?t=lt0eoc

46 Schneider, H. and Saphir, A., (2021) Fed's Powell holds fast to 'this year' timeline for bond-buying taper, *Reuters*, https://www.reuters.com/business/feds-powell-weighs-delta-risk-gives-no-signal-start-bond-buying-taper-2021-08-27/

47 Patteson, C. (2021) Jerome Powell suggests it's time to stop describing inflation as 'transitory', NY Post, https://nypost.com/2021/11/30/jerome-powell-its-time-to-retire-term-transitory-inflation/

48 Patton, M. (2021) Inflation Surge To Continue: Here Are 3 Reasons Why, *Forbes*, https://www.forbes.com/sites/mikepatton/2021/08/18/inflation-surge-to-continue-here-are-3-reasons-why/?sh=53ff9abc690a

49 St. Louis Fed https://alfred.stlouisfed.org/series?seid=FYFSD

- The Federal Reserve added $1.8 billion to their balance sheet in a five month period during 2020.[50]
- By the summer of 2021, "Firms were detailing the persistent nature of the disruptions in their supply chains. Labor shortages were multiplying, adding to the cost-push drivers of inflation. Few, if any, companies expected these two issues to be resolved any time soon – and said so on one earnings call after another."[51]

It was obvious to many commentators at the time that inflation would not be transitory. Mohamed El-Erian said that "The characterization of inflation as transitory is probably the worst inflation call in the history of the Federal Reserve, and it results in a high probability of a policy mistake."[52] He also said "But by the end of the summer, it became clear that behavior on the ground was changing, especially as inflation continued its steady ascent (to 6.2% for the headline consumer price index in October and 4.1% for the core personal consumption expenditures price index, the Fed's preferred gauge). Yet, consistent with classic behavioral traps, the Fed remained wedded to a transitory concept, with Chair Jerome Powell insisting again in the last week of November that 'inflation will move down significantly over the next year'."[53]

# Missing the big Picture

## Cancer screenings

There is often a question of who should get screened for various cancers. Often, the guidance is based on how old someone is and that person's family history. When doctors and scientist set those guidelines, they consider the rates of true positives, false positives, true negatives, and false negatives. That may not be enough.

One book notes that in addition to concerns about false positives and false negatives, we should also be concerned about "overdiagnosis, in which a slow-growing cancer found by screening never would have caused harm or required treatment during a patient's lifetime. Because of overdiagnosis, the number of cancers found at an earlier stage is also an inaccurate measure of whether a screening test can save lives."[54] In other words, the number of positive diagnosis includes many cases where the cancer would never have been harmful. Detecting these cases, with the cost of screening and subsequent life disruption, is actually worse than not screening for them.

This article mentions that cases detected through screening have higher five year survival rates then after screening than cases diagnosed after symptoms have shown up. The article states that this is partly because in cases detected through screening, the cancer has not progressed as far and partly because many cases captured in screening will never cause symptoms or health complications.  The article provides the figure below which illustrates the first reason.

[50] Steil, B. and Rocca, B. (2020) Why the Fed Bond Binge Will Boost Inflation, *Council on Foreign Relations*. https://www.cfr.org/blog/why-fed-bond-binge-will-boost-inflation-1

[51] El-Erian, M. Can the Fed Overcome Its Transitory Policy Mistake? *Project Syndicate*. https://www.project-syndicate.org/commentary/us-federal-reserve-transitory-inflation-trap-by-mohamed-a-el-erian-2021-12

[52] Cox, J. El-Erian (2021) says 'transitory' was the 'worst inflation call in the history' of the Fed. *CNBC*. https://www.cnbc.com/2021/12/13/el-erian-says-transitory-was-the-worst-inflation-call-in-the-history-of-the-fed.html

[53] El-Erian, (2021). Can the Fed Overcome Its Transitory Policy Mistake? Project-Syndicate. https://www.project-syndicate.org/commentary/us-federal-reserve-transitory-inflation-trap-by-mohamed-a-el-erian-2021-12

[54] National Cancer Institute (2018) Crunching Numbers: What Cancer Screening Statistics Really Tell Us, https://www.cancer.gov/about-cancer/screening/research/what-screening-statistics-mean

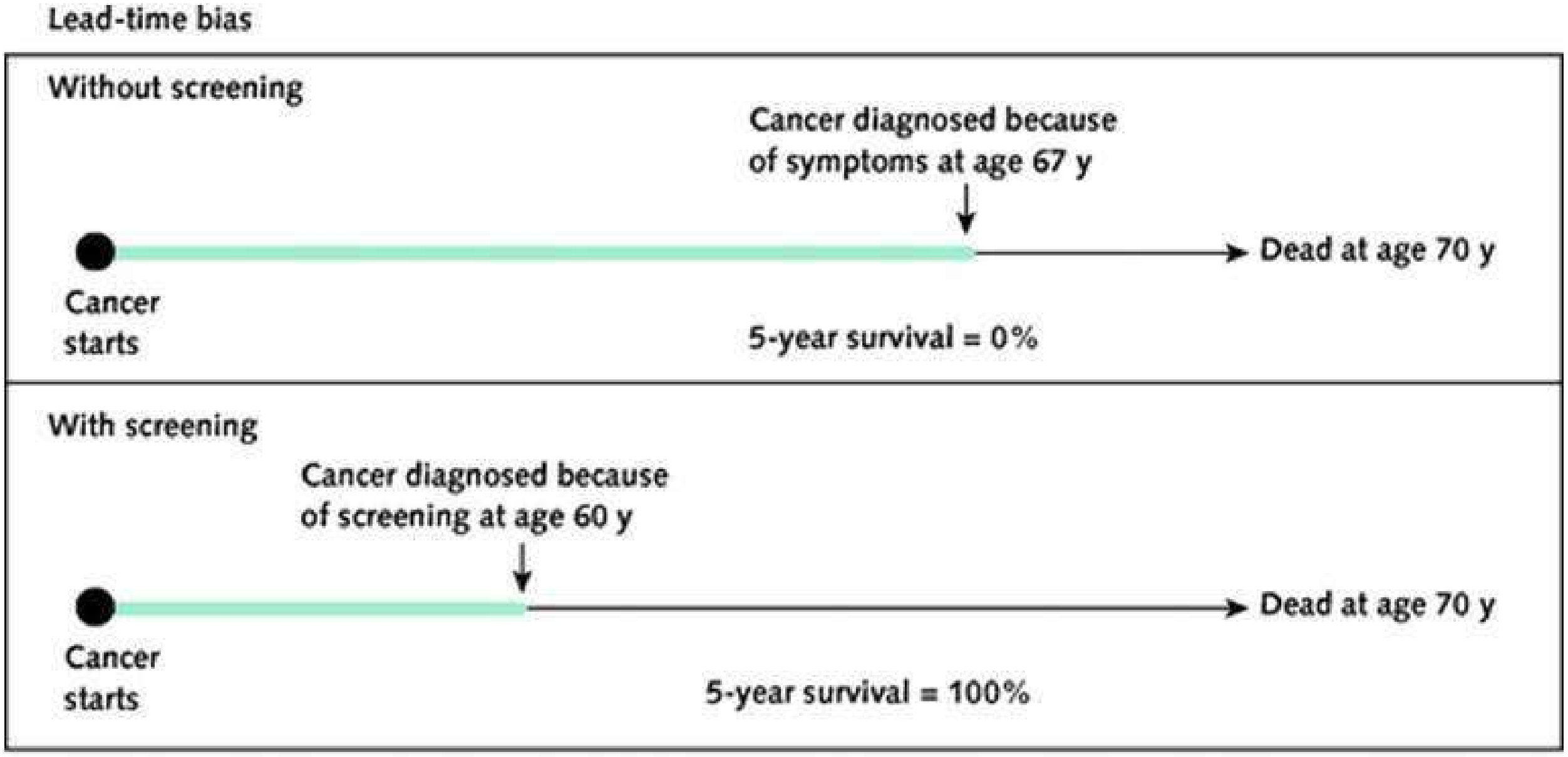

## Recommendation

Most actions have risks and costs, including opportunity costs – the inability to do something else because we are doing this.  We need to understand these risks and costs as well as the possible benefits and evaluate the corresponding trade-offs.  We can then make more complete recommendations.

# Not Sharing Data and Models

Many models are proprietary and secret.  There are good arguments for this in pricing insurance and underwriting loans where companies are competing in the market for customers.  It's a lot harder to make the case for secrecy when the government is using models.

## COMPAS

The Correctional Offender Management Profiling for Alternative Sanctions tool[55], commonly known as COMPAS, is used by "criminal justice agencies across the nation to inform decisions regarding the placement, supervision and case management of offenders."

### Child welfare tools

[55] Equivant, (2019) Practitioner's Guide to COMPAS Core. https://www.equivant.com/wp-content/uploads/Practitioners-Guide-to-COMPAS-Core-040419.pdf

In some jurisdictions, the government is using AI tools to determine when children might be at risk of harm[56]. In some cases, this results in the government removing the child from their parents.

## Recommendation

In both of these cases, the government is using these models to affect people's basic rights – freedom and family integrity. While it is good to have an objective process, the AI tools should not be a black box. Affected parties should know enough to challenge the model results and to change their behavior to improve their scores.

# Conclusion

With big data and computational power we could only dream of twenty years ago, we now have the power to test lots of things. Looking at lots of predictors, we are bound to find something that stands out. This makes it important that we are rigorous with our analysis. After any analysis, we should ask ourselves "Would I be convinced if our data showed the opposite of what we expected?", "What else could be causing our observed results?", and "How significant, material, and compelling are these results?".

As we think about this, we should:

- Ensure that we have the right target that is consistent with how we will use the model.
- Consider whether correlated variables might be causing the effect that we are seeing.
- Share our model and underlying data, to the extent that we can.
- Partner with researchers (or peer reviewers) who have different views and expectations than we do[57].
- Separate our objective view of what the data is telling us from our recommendations on how we should act.

  These principles should help us preserve our credibility and our seat the table for these conversations.

[56] Ho, S., Burke, G., Not magic: Opaque AI tool may flag parents with disabilities, *SF Gate*, March 14, 2023, https://web.archive.org/web/20230316011738/https://www.sfgate.com/news/politics/article/not-magic-opaque-ai-tool-may-flag-parents-with-17840002.php

[57] Knott, K. (2023) Research Finds No Gender Bias in Academic Science. *Inside Higher Ed*. https://www.insidehighered.com/news/faculty-issues/diversity-equity/2023/04/27/research-finds-no-gender-bias-academic-science and Ceci, S. J., Kahn, S., & Williams, W. M. (2023). Exploring Gender Bias in Six Key Domains of Academic Science: An Adversarial Collaboration. *Psychological Science in the Public Interest.* https://doi.org/10.1177/15291006231163179